\documentclass[epj]{svjour}
\usepackage[utf8]{inputenc}
\usepackage{amsmath}
\usepackage{amsfonts}
\usepackage{amssymb}
\usepackage{braket}
\usepackage{bbm}
\usepackage{graphicx}
\usepackage{tikz}
\usepackage{changepage}
\usepackage{multirow}
\usepackage{mathtools}

\def \l( {\left(}
\def \r) {\right)}
\def \b {\hat{b}}
\def \bplus {\hat{b}^\dagger}
\def \Z {\mathcal{Z}}
\def \D {\mathcal{D}}
\def \C {\mathcal{C}}

\def \n {\hat{n}}
\def \e {\epsilon}
\def \O {\mathcal{O}}

\def \Utilde {\tilde{U}}
\def \mutilde {\tilde{\mu}}
\def \Vtilde {\tilde{V}}
\begin{document}
\title{Haldane Insulator in the 1D Nearest-Neighbor Extended Bose-Hubbard Model with Cavity-Mediated Long-Range Interactions}
\author{Johannes Sicks\inst{1} \and Heiko Rieger\inst{1}
}                     
%
%
\institute{Theoretical Physics, Saarland University, Campus E2.6, 66123 Saarbr{\"u}cken, Germany}
\date{Received: date / Revised version: date}
%
\abstract{
In the one-dimensional Bose-Hubbard model with on-site and nearest neighbor interactions, a gapped phase characterized by an exotic non-local order parameter emerges, the Haldane insulator. Bose-Hubbard models with cavity-mediated global range interactions display phase diagrams, which are very similar to those with nearest neighbor repulsive interactions, but the Haldane phase remains elusive there. Here we study the one-dimensional Bose-Hubbard model with nearest-neighbor and cavity-mediated global-range interactions and scrutinize the existence of a Haldane Insulator phase. With the help of extensive quantum Monte-Carlo simulations we find that in the Bose-Hubbard model with only cavity-mediated global-range interactions no Haldane phase exists. For a combination of both interactions, the Haldane Insulator phase shrinks rapidly with increasing strength of the cavity-mediated global-range interactions. Thus, in spite of the otherwise very similar behavior the mean-field like cavity-mediated interactions strongly suppress the non-local order favored by nearest neighbor repulsion in some regions of the phase diagram.
%
} 
\maketitle
\section{Introduction\label{sec:Introduction}}
For several decades, the Bose-Hubbard model (BHM) \cite{Fisher1989} has attracted continued interest. In its most simplistic form, it exhibits two phases in the ground state: A Mott insulator (MI) phase and a superfluid (SF) phase, depending if on-site repulsion or nearest-neighbor hopping dominates.
Through the years, quantum Monte-Carlo (QMC) methods contributed greatly to the investigation of quantum critical phenomena. Here, one must especially emphasize path-integral Monte-Carlo \cite{Pollock1987,Ceperley1995}, world-line QMC \cite{Batrouni1992,Batrouni1995}, worm-algorithm QMC \cite{Prokofev1998,Prokofev1998_aug,Prokofev2001,Prokofev2004} and, as a derivative, the stochastic Green's function algorithm \cite{Rousseau2008_may,Rousseau2008_nov}.

Also approximate techniques were applied to the BHM, like mean-field theory \cite{Fisher1989,vanOosten2001,Dhar2011}, strong coupling expansion \cite{Elstner1999}, Gutzwiller wave function variational calculation \cite{Rokhsar1991,Krauth1992} and density matrix renormalisation group method \cite{Kuehner1998}.

First experimental realizations of the BHM involved ultracold bosons trapped in optical lattices \cite{Jaksch1998,Greiner2002} and initiated studies of many-body bosonic gases with additional potentials and interactions \cite{Kim2004,Landig2016,Hruby2018}. These extended models generally feature new phases \cite{Batrouni2006,Capogrosso-Sansone2007,Capogrosso-Sansone2008,Capogrosso-Sansone2010,Ohgoe2012_aug,Batrouni2013}.

Analyzing extended models, several inclusions to the BHM were made, e.g. the addition of harmonic confining potentials \cite{Rigol2009,Kato2009}, three-body interactions \cite{Greschner2013}, disordered potentials \cite{Gurarie2009,Niederle2016,Zhang2019}, long-range dipolar interactions \cite{Capogrosso-Sansone2010,Zhang2015}, nearest-neighbor interactions \cite{Kuehner1998,Batrouni2006,Ohgoe2012_aug,Batrouni2013,Sengupta2005,Iskin2011,Kimura2011,Rossini2012,Kawaki2017}, next-nearest-neighbor interactions \cite{Batrouni1995,Herbert2001,Schmid2004,DallaTorre2006}, next-nearest-\linebreak neighbor hopping \cite{Chen2008}, cavity-mediated long-range interactions \cite{Hruby2018,Dogra2016,Flottat2017} and a combination of nearest-neighbor and long-range interactions \cite{Landig2016,Zhang2019,Bogner2019}.

For the nearest-neighbor (NN) as well as the cavity-mediated long-range (LR)  interaction the extended BHM exhibits additional phases: the density wave (DW) phase and the supersolid (SS) phase. Furthermore it was shown, that for the 1D NN extended BHM a Haldane insulator (HI) phase exists \cite{Batrouni2013,DallaTorre2006,Berg2008}.
Originally, the HI was introduced for the Spin-$S$ antiferromagnetic Heisenberg chain, where the ground state has an excitation gap when $S$ is integer and no gap when $S$ is half integer \cite{Haldane1983,Haldane1983_2}. A reduced BHM, where site occupation numbers are restricted to $0, 1$ and $2$, can be mapped onto the Spin-$1$ Heisenberg chain \cite{DallaTorre2006,Berg2008}. For density $\rho=1$, the deviation of the site occupation numbers from $1$, $\delta \hat{n}_i = \hat{n}_i - \rho$, corresponds to the $\hat{S}^z$ operator in the Spin-$1$ Heisenberg chain.

An interesting question is, whether the HI phase can also occur in the BHM with cavity-mediated interactions, since on the mean-field level it is equivalent to the BHM with NN interactions \cite{Dogra2016}. In this paper we will address this question with the exact QMC worm-algorithm \cite{Prokofev1998,Prokofev1998_aug}.

The paper is organized as follows: In Sec. \ref{sec:Model} we introduce the BHM with extended NN and LR interactions and show that both additional interactions lead to similar terms in the mean-field approximation. Sec. \ref{sec:Worm-Algorithm} outlines the QMC worm-algorithm. The measured observables and results for different parameter settings are discussed in Sec. \ref{sec:Results}. We account chain lengths up to $L=192$.

\section{Model\label{sec:Model}}
We consider the one-dimensional extended Bose-Hubbard model with nearest-neighbor short-range and cavity-\linebreak mediated long-range interactions (NNLR-BHM) in the \linebreak grand-canonical ensemble. It is defined by the Hamiltonian
\begin{align}
\mathcal{\hat{H}} = &-t \sum_{\langle i,j \rangle} \l( \bplus_i \b_j + h.c. \r) + \frac{U}{2}\sum_{i} \n_i (\n_i - 1) \nonumber\\
&+ V \sum_{\langle i,j \rangle} \n_i\n_j - \mu \sum_{i}\n_i \nonumber \\
&-\dfrac{U_d}{L}\l( \sum_{e} \n_e - \sum_{o} \n_o \r) ^2 ~.
\label{eq:Hamiltonian}
\end{align}
The first term describes the nearest-neighbor hopping between two sites with the hopping strength $t$. $\bplus_i~(\b_i)$ are the bosonic creation (annihilation) operators. The second term describes the on-site repulsion $(U>0)$ and $\n_i$ is the number operator. The third term defines a nearest-neighbor repulsion $(V>0)$. The fourth term contains the chemical potential $\mu$. We use the grand-canonical ensemble, thus keeping $\mu$ fixed and let the total particle number $N$ vary. The cavity-mediated interaction is introduced in the last term. $U_d$ is the misbalance parameter, $L$ the chain length and $\sum_{e(o)}$ the sum over all even (odd) sites. We apply periodic boundary conditions.
The total misbalance is defined as
\begin{equation}
\D = \l( \sum_{e} \n_e - \sum_{o} \n_o \r) ,
\end{equation}
where $\D$ ranges between $-N$ (all bosons on odd sites) and $+N$ (all bosons on even sites).

In Dogra \textit{et al.}  \cite{Dogra2016} it was shown that the mean-field (MF) expressions of the LR and NN interactions are equivalent. In mean-field approximation the decoupled expressions for kinetic energy, NN and LR interactions read
\begin{align}
\bplus_i \b_j &\approx \langle \bplus_i \rangle \b_j + \bplus_i \langle \b_j \rangle  - \langle \bplus_i \rangle \langle \b_j \rangle = \psi \l( \bplus_i + \b_j \r) - \psi^2 , \nonumber \\
\sum_{\langle i,j \rangle}\n_i \n_j &\approx \dfrac{z}{2} \sum_i {\n_i}^2-z~\vartheta~ \D + \dfrac{z}{4}~L~\vartheta^2, \nonumber \\
\D^2 & \approx 2\langle \D \rangle \D - \langle \D \rangle^2 = L~ \theta~ \D - \dfrac{L^2}{4}\theta^2 ,
\label{eq:MF_decoupling}
\end{align}
with the coordination number $z$ and the order parameters $\psi = \langle \bplus_i \rangle = \langle \b_j \rangle$, $\vartheta = \langle \n_i \rangle - \langle \n_j \rangle$ and $\theta = 2\langle \D \rangle / L$.\\
Hence, on a \textit{mean-field level} the LR-BHM term in (\ref{eq:Hamiltonian}) is equivalent to the NN-BHM term:
\begin{align}
\mathcal{\tilde{\hat{H}}} = &-t \sum_{\langle i,j \rangle} \l( \bplus_i \b_j + h.c. \r) + \frac{\Utilde}{2}\sum_{i} \n_i (\n_i - 1) \nonumber\\
&+ \Vtilde \sum_{\langle i,j \rangle} \n_i\n_j - \mutilde \sum_{i}\n_i ~,
\label{eq:MF_Hamiltonian}
\end{align}
where $\Utilde = U-U_d$, $\mutilde = \mu + U_d/2$ and $\Vtilde = V + U_d/z$.

In (\ref{eq:MF_Hamiltonian}) the transformed LR interaction increases the NN interaction, as intuitively expected. Furthermore, the chemical potential is shifted, such that even for negative values of $\mu$ the system can localize in a DW or MI phase and the rescaling of the on-site potential leads to a narrower lobe width. 

Ultimately, since the HI phase was already found in Hamiltonian (\ref{eq:Hamiltonian}) with $U_d=0$ \cite{Batrouni2013,Rossini2012,DallaTorre2006,Berg2008} and since on the MF level NN and LR interactions are equivalent one is lead to ask whether the HI phase exists also for $U_d>0$. This is the question that we will answer in the following using a QMC algorithm.\\

\section{Worm-Algorithm\label{sec:Worm-Algorithm}}
To determine the ground state properties of the Hamiltonian (\ref{eq:Hamiltonian}) we use the quantum Monte-Carlo (QMC) worm-algorithm \cite{Prokofev1998,Prokofev1998_aug,Pollet2007}. It relies on the Dyson series 
where the $d$-dimensional quantum system is mapped onto a ($d+1$)-dimensional classical one.
The partition function  reads
\begin{align}
\Z = &\sum_{m=0}^{\infty} \sum_{\textbf{n}_1 \dots \textbf{n}_m} e^{-\beta \epsilon_1} \int_0^\beta d\tau_m \cdots \int_{0}^{\tau_{2}} d\tau_1 \\ \nonumber
&\times\l( e^{\tau_m \e_1} \hat{V}_{\textbf{n}_1 \textbf{n}_m} e^{-\tau_m \e_m} \r) \cdots \l( e^{\tau_1 \e_2} \hat{V}_{\textbf{n}_2 \textbf{n}_1} e^{-\tau_1 \e_1} \r) ,
\label{eq:Z}
\end{align}
with the inverse temperature $\beta$ and $\hat{V}_{\textbf{n}_i \textbf{n}_j}= \braket{\textbf{n}_i | \hat{V} | \textbf{n}_j}$. The off-diagonal part of the Hamiltonian (\ref{eq:Hamiltonian}) is \linebreak$\hat{V} = t \sum_{\langle i,j \rangle} ( \bplus_i \b_j + h.c. ) $, $\ket{\textbf{n}_i}$ are the Fock states of the diagonal Hamiltonian and $\e_i$ are the diagonal energy values of the respective Fock states.

In the worm-algorithm the configuration space is expanded by a creator and annihilator pair in the form $\b_i(\tau') = ( e^{\tau' \e'_a}~ \b_{i}~ e^{-\tau' \e_a}) $ and vice versa for $\bplus_i(\tau')$. One operator is assigned as the worm head and the other one as the worm tail. The head moves through the given state and can create, delete or relink vertices, where bosons hop from one site to a neighboring one. When it reaches the tail the worm gets deleted.

Between a head and tail, the total particle number will be increased or decreased in comparison to the initial state. Thus the worm-algorithm performs grand-canonical update steps. When the update procedure is complete and the worm deleted, one can calculate canonical observables 
by importance sampling
\begin{equation}
\langle \O \rangle = \frac{1}{\Z} \sum_{\C} \O(\C)\Z(\C) ~,
\end{equation}
with $\C$ denoting states with fixed total particle numbers.

We consider chain lengths up to $L=192$ and an on-site repulsion of $U=1$. The inverse temperature was set to $\beta = 128$ as comparisons with lower temperatures showed no sufficient difference for the obtained order parameters.
\\
\section{Results \label{sec:Results}}
\subsection{Measured Observables \label{subsec:MeasuredObservables}}
In the NNLR-BHM various observables are of interest. The particle density is $\rho = \sum_i \langle \n_i \rangle / L$ and the superfluid density can be calculated via \cite{Pollock1987,Ceperley1995}
\begin{equation}
\rho_s = \dfrac{\langle W^2 \rangle L}{2t\beta} ~,
\end{equation}
where $W$ is the winding number, which is the difference of bosons crossing over one side of the periodic boundary conditions minus the crossing over the other side. The density-density correlation and structure factor are defined as
\begin{align}
D(r) &= \dfrac{1}{L} \sum_i \langle \n_i \n_{i+r} \rangle ~,\\
S(k) &= \frac{1}{L} \sum_r e^{\imath k r} D(r) ~.
\end{align}
For $k=\pi$ the structure factor gives the misbalance order parameter $\theta = 2\langle \D \rangle/L$. 

These order parameters would be sufficient to distinguish between the Mott insulator (MI) phase, superfluid (SF) phase, density wave (DW) phase and supersolid (SS) phase.

For a fixed density $\rho=1$ and large $U$ and $V$ values, the  site occupation is practically restricted to $0, 1$ and $2$. As a result, the difference between particle number and density $\delta \n_i = \n_i - \rho$ has the same Eigenvalues as the $\hat{S}^z$ spin operator from the Spin-$1$ Heisenberg chain ($-1,0,1$) and thus both models are similar.

It was shown with a non-local unitary transformation of the Heisenberg chain, that the HI breaks a hidden $\mathbb{Z}_2$ symmetry \cite{Kennedy1992}. To determine the HI we introduce two non-local observables, the string and the parity operators \cite{Berg2008}:
\begin{align}
\O_s \l( |i-j|\r) &= \left\langle \delta \n_i \exp \left\lbrace \imath \pi \sum_{k=i}^{j} \delta \n_k \right\rbrace \delta \n_j \right\rangle ~, \\
\O_p \l( |i-j|\r) &= \left\langle \exp \left\lbrace \imath \pi \sum_{k=i}^{j} \delta \n_k \right\rbrace \right\rangle ~.
\end{align}
We evaluate both observables for $|i-j| = L/2$.

In Table \ref{tab:transitions}, all possible phases with $\rho=1$ are shown together with their order parameter values. The HI can be described as a charged ordered state 
\begin{equation}
\dots 0,+1,0,\dots,0,-1,0,\dots,0,+1,0,\dots,0,-1,0,\dots
\label{eq:charge_ordered_state}
\end{equation}
while the numbers represent $\delta\n_i$ with an undetermined amount of $0$ between each $+1$ and $-1$. On the other hand, the MI is a dilute gas of particle-hole pairs, thus no global ordering emerges but local fluctuations which can result in forms like
\begin{equation}
\dots 0,\underbracket{+1,0,-1},\underbracket{-1,+1},0,0,\underbracket{+1,0,0,-1},\underbracket{+1,0,-1},0,\dots ,
\label{eq:dilute_gas}
 \end{equation} 
where two consecutive $-1$ or $+1$ emerge and counteract the global ordering \cite{Berg2008}. Therefore in finite systems, it is possible that the particles $(+1)$ and holes $(-1)$ wind around the chain, which results in a non-zero superfluid density. However, in the limit ${L \rightarrow \infty}$, $\rho_s$ disappears for the HI and MI phases.
\begin{table}[htbp] 
\centering
\begin{tabular}{|l|c c c c|}
\hline
& $\rho_s$ & $S(\pi),\theta$ & $\O_s(L/2)$ & $\O_p(L/2)$ \\  \hline 
SF & $\neq 0$ & $0$ & $0$ & $0$ \\ 
SS & $\neq 0$ & $\neq 0$ & $\neq 0$ & $\neq 0$ \\ 
DW & $0$ & $\neq 0$ & $\neq 0$ & $\neq 0$ \\ 
MI & $0^*$ & $0$ & $0$ & $\neq 0$ \\ 
HI & $0^*$ & $0$ & $\neq 0$ & $0$ \\ 
\hline
\end{tabular}
\caption{Order parameters at $\rho=1$ for different phases. (*) In finite chain lengths, the superfluid density can attain non-zero values.}\label{tab:transitions}
\end{table}

We consider three different parameter regimes of the NNLR-BHM. In the first part, we set the NN interaction to $V=0.75$ and the LR interaction to $U_d = 0$ and compare our results with \cite{Batrouni2013}. In the second part we fix $V=0$ and $U_d = 0.6$. Finally, we include NN as well as LR interactions and compare the behavior of obtained phases to the previous cases. Therefore, we set $V=0.75$ and increase $U_d$ in size.

For all cases, we focus on the $\rho=1$ lobe of the \linebreak ${\mu/U-t/U}$ diagram and determine the phases for commensurate filling to see whether a HI phase occurs. So, we tune the chemical potential carefully such that the particle density $\rho$ equals $1$ and then perform measurements of the order parameters.
\subsection{Results for the NN-BHM \label{subsec:ResultsfortheNN-BHM}}
In this subsection, we neglect the LR interaction, i.e. $(U_d=0)$ and consider only the NN extended BHM with $V=0.75$. Since $zV > U$ the DW(2,0) phase appears instead of the MI(1) phase. Furthermore, calculations for the $t=0$ case yield that the DW(2,0) phase occures for ${1 < \mu/U < 2}$.

\begin{figure}\centering
\includegraphics[width=\columnwidth]{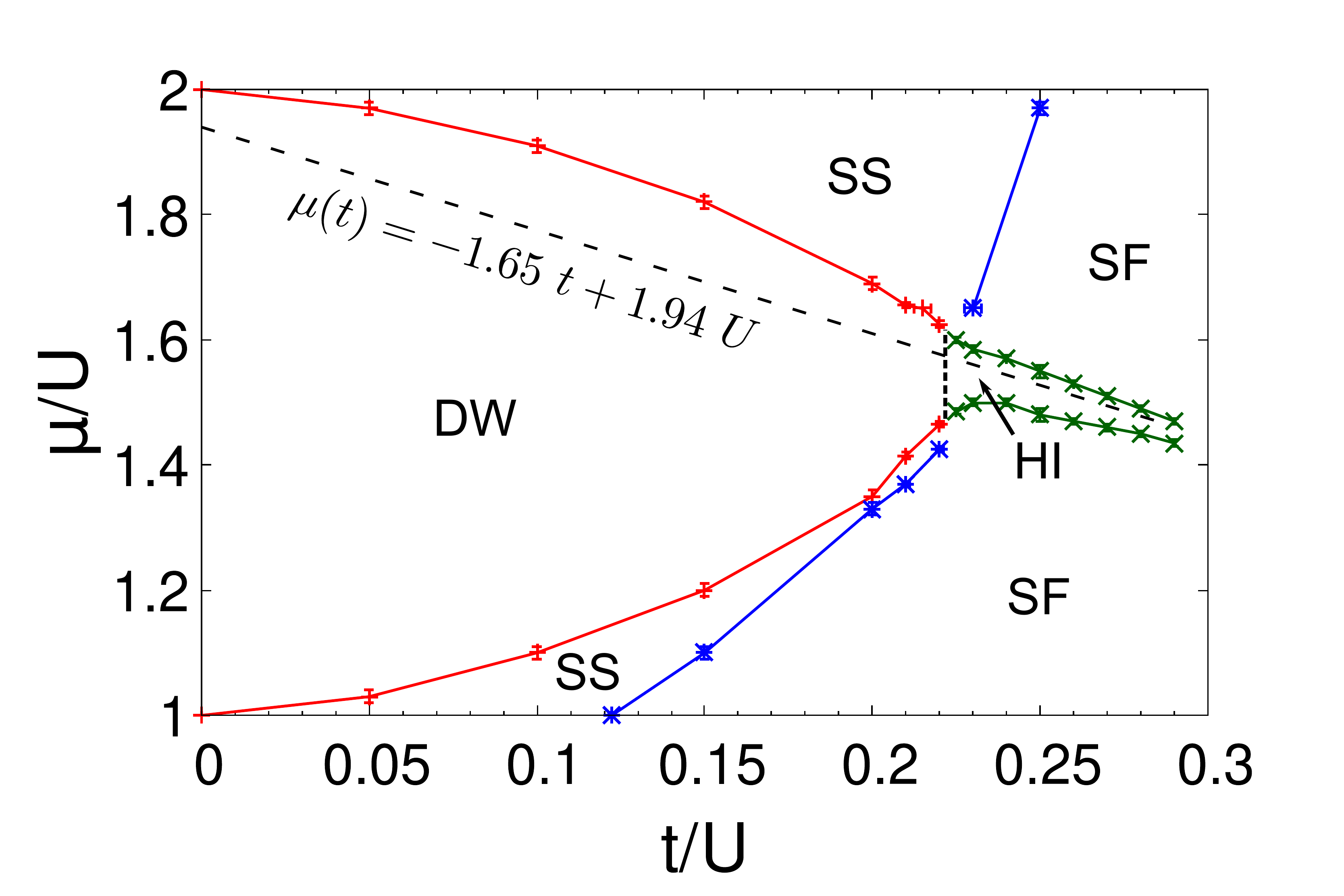}%
\caption{\label{fig:phase_diagram_Ud_0} Phase diagram of the NN-BHM with $V=0.75$. Red line: Transition between DW(2,0) and SS phases. Green line: Transition between HI and SF phases. Blue line: Transition between SF and SS phases. The chemical potentials used in Fig. \ref{fig:Ud_0} are obtained from the linear function $\mu(t)=-1.65~t+1.94~U$.}
\end{figure}

The phase diagram for the $\rho = 1$ lobe is depicted in Fig. \ref{fig:phase_diagram_Ud_0}. There are four different phases: The DW(2,0) phase in the red lobe, surrounded by a SS phase up to the blue line where the transition to the SF phase occurs. Between the green lines, the HI phase is present and transits into the DW phase at around $t=0.22$. Our results are in good agreement with \cite{Batrouni2013}, where the 1D phase diagram with NN potential was calculated with density matrix renormalization group and stochastic Green's function methods.
 
\begin{figure}\centering
\includegraphics[width=\columnwidth]{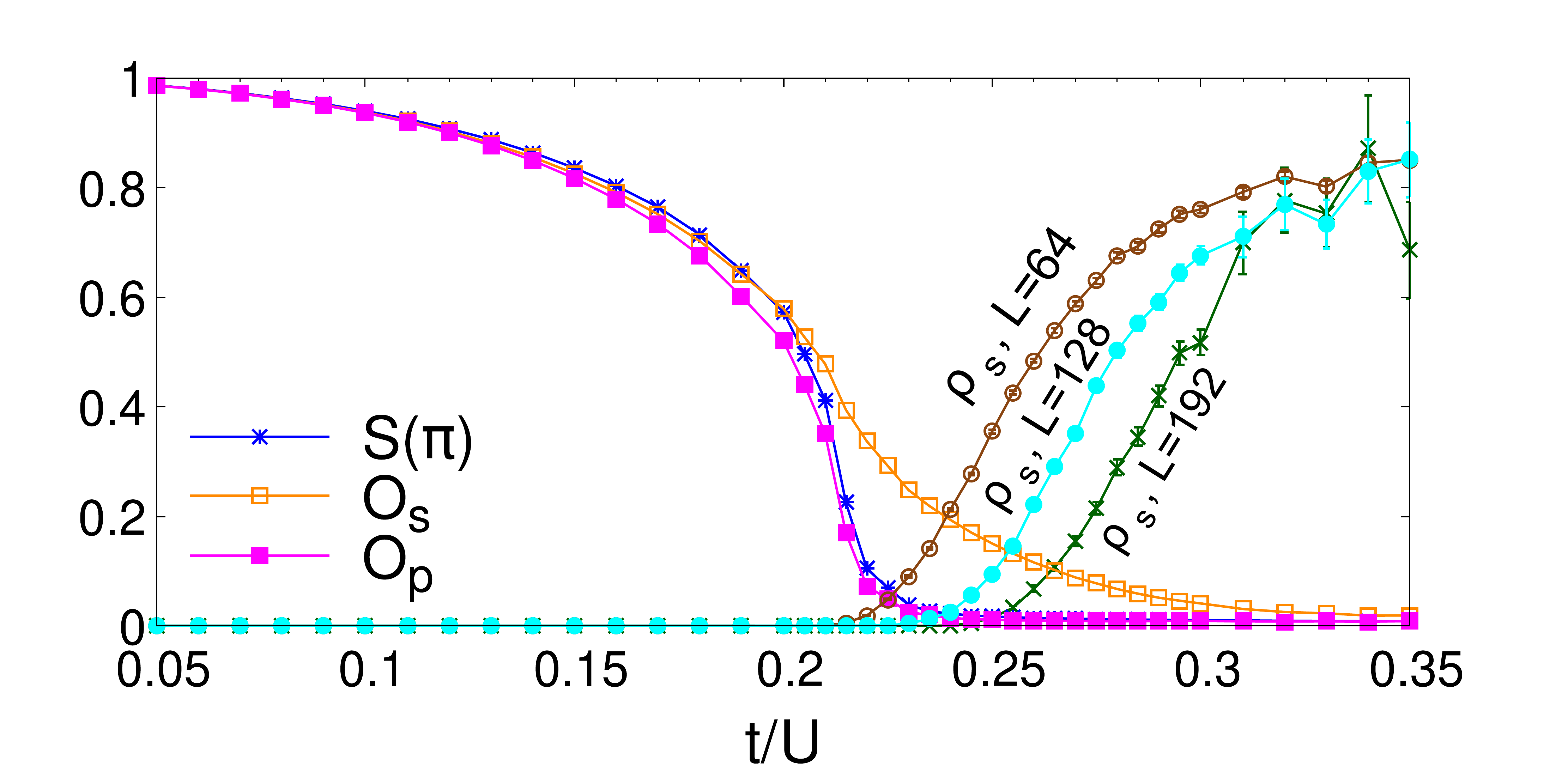}%
\caption{\label{fig:Ud_0} Order parameters as a function of $t/U$ for $\mu(t)=-1.65~t+1.94~U$ and $V/U=0.75$ in the NN extended BHM at $\rho=1$. Structure factor $S(\pi)$, string $\O_s$ order parameter and parity $\O_p$ order parameter are given for $L=192$. The superfluid density is shown for $L=64,128,192$. DW phase: $\rho_s = 0$ and $S(\pi)\neq0, \O_s\neq0 ,\O_p\neq0$. HI phase: $\rho_s \rightarrow 0$ for increasing chain lengths, $S(\pi) = \O_p = 0$ and $\O_s \neq 0$. SF phase: $\rho_s \neq 0$ and $S(\pi) = \O_s = \O_p = 0$. The phase transition from DW phase to HI phase is at around $t=0.22$ while the HI-SF transition lies between $0.32<t<0.33$.}
\end{figure}

Next, we fix $\mu$ to a linear equation depending on $t$, like depicted in Fig. \ref{fig:phase_diagram_Ud_0} and increase $t$ from $0$ to $0.35$, thus traversing the whole $\rho=1$ lobe to the tip and into the HI phase. The results for the order parameters is shown in Fig \ref{fig:Ud_0}.
Here, the DW phase, in which $\rho_s$ is zero and all other parameters are non-zero, persists up to $t \approx 0.22$ where the transition to the HI phase occurs. There, $S(\pi)$ and $\O_p$ drop to zero while $\O_s$ decays but stays finite. The superfluid density is non-zero but size dependent and vanishes for larger system sizes until the transition to the SF phase occurs. For the SF phase all order parameters are zero except the superfluid density.

\begin{figure}\centering
\includegraphics[width=.7\columnwidth]{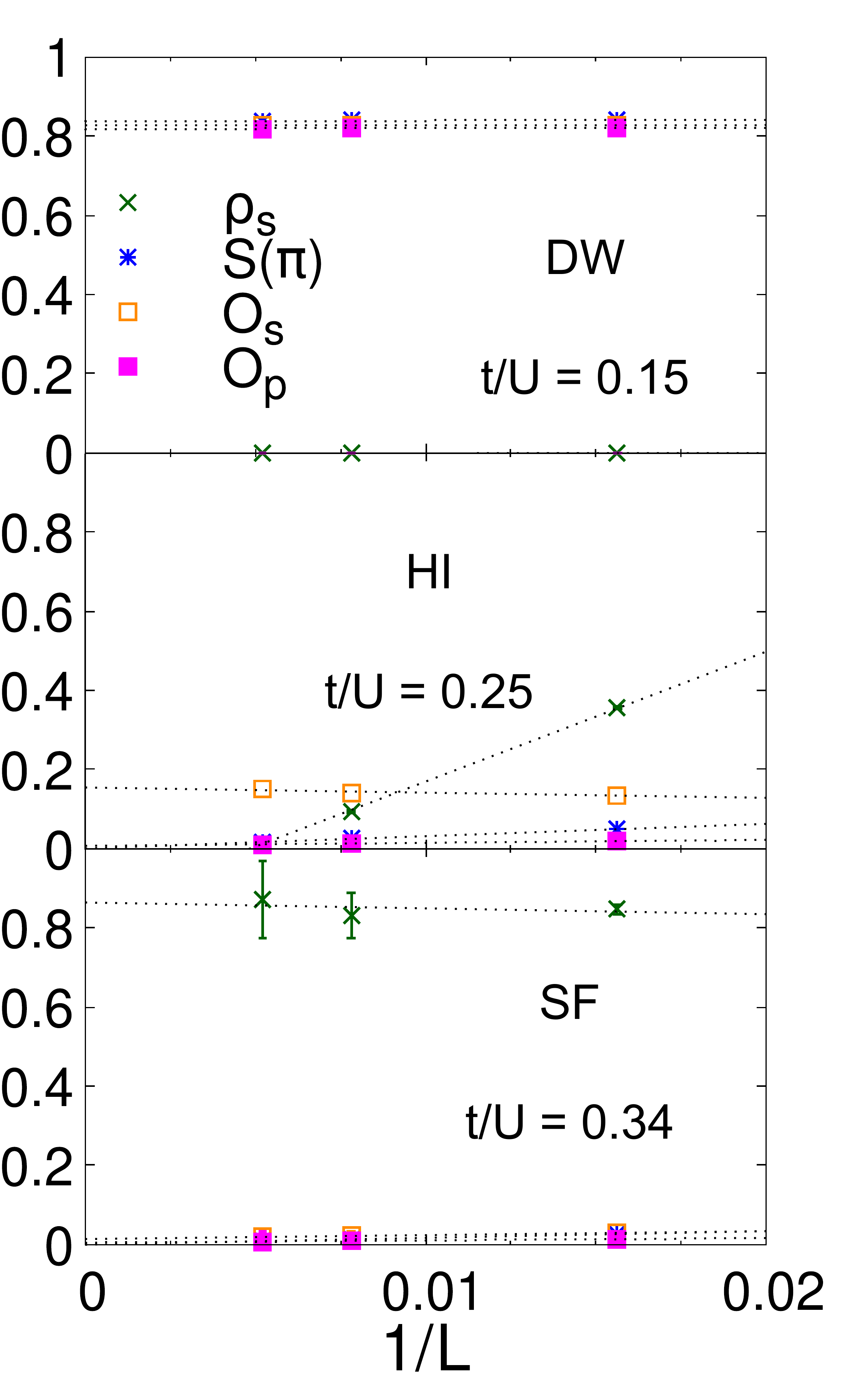}%
\caption{\label{fig:L_scaling_merged} Finite size dependence of the order parameters. Top: DW phase where $S(\pi)$, $\O_s$ and $\O_p$ have finite values and $\rho_s$ vanishes. All parameters are practically independent of $L$. Middle: HI phase, $S(\pi)$ and $\O_p$ vanish for $L\rightarrow \infty$ and $\O_s \neq 0$. The superfluid density approaches zero at a finite value of $L$ (see text for further details). Bottom: SF phase where only $\rho_s$ persists for infinite $L$ values and $S(\pi)$, $\O_s $, $\O_p$ approaching zero for infinite sizes.}
\end{figure}

Next, we study the finite size behavior of the order parameters in the different phases. In Fig. \ref{fig:L_scaling_merged} three generic points extracted from Fig. \ref{fig:Ud_0} are shown. On the top the DW phase for $t=0.15$, in the middle the HI phase for $t=0.22$ and on the bottom the SF phase for $t=0.35$.

For the DW phase all order parameters stay constant. As expected $S(\pi)$, $\O_s$ and $\O_p$ attain finite values while $\rho_s = 0$ for all chain lengths.
In the HI phase $S(\pi)$ and $\O_p$ are finite for small sizes and become zero for infinite $L$. The string order parameter persists for larger sizes and approaches a finite value. The superfluid density decreases for increasing lengths and vanishes completely for a large (but finite) $L$.

This behavior underlines the breaking of the hidden $\mathbb{Z}_2$ symmetry. For small sizes the particles and holes can potentially wind around the chain, leading to a finite superfluid density value. This fluctuation is dependent on the amount of particles and holes in the system, meaning when there are a lot - like close to the DW phase - the fluctuations get small and when there are only a few, the fluctuations get high. When $L$ is smaller than this fluctuation length, winding happens and the superfluid density is greater than zero. Otherwise, there exists a finite chain length where no winding and thus no superfluid density exists any more.

The order parameters in the SF phase behave opposite to the DW phase. The superfluid density is non-zero and does not vanish for infinite sizes while $S(\pi)$, $\O_s$ and $\O_p$ are very small for tiny lengths and become zero for $L\rightarrow\infty$.

We can visualize the finite size effects in the HI phase also by the Green's function
\begin{equation}
G(r) = \dfrac{1}{2L} \sum_i \left\langle \bplus_i \b_{i+r} + h.c. \right\rangle .
\end{equation}
The worm algorithm can directly calculate the Green's function, which is a degree of spatial movement in the system and thus correlated to the winding number \cite{Ceperley1995}.
\begin{figure}\centering
\includegraphics[width=\columnwidth]{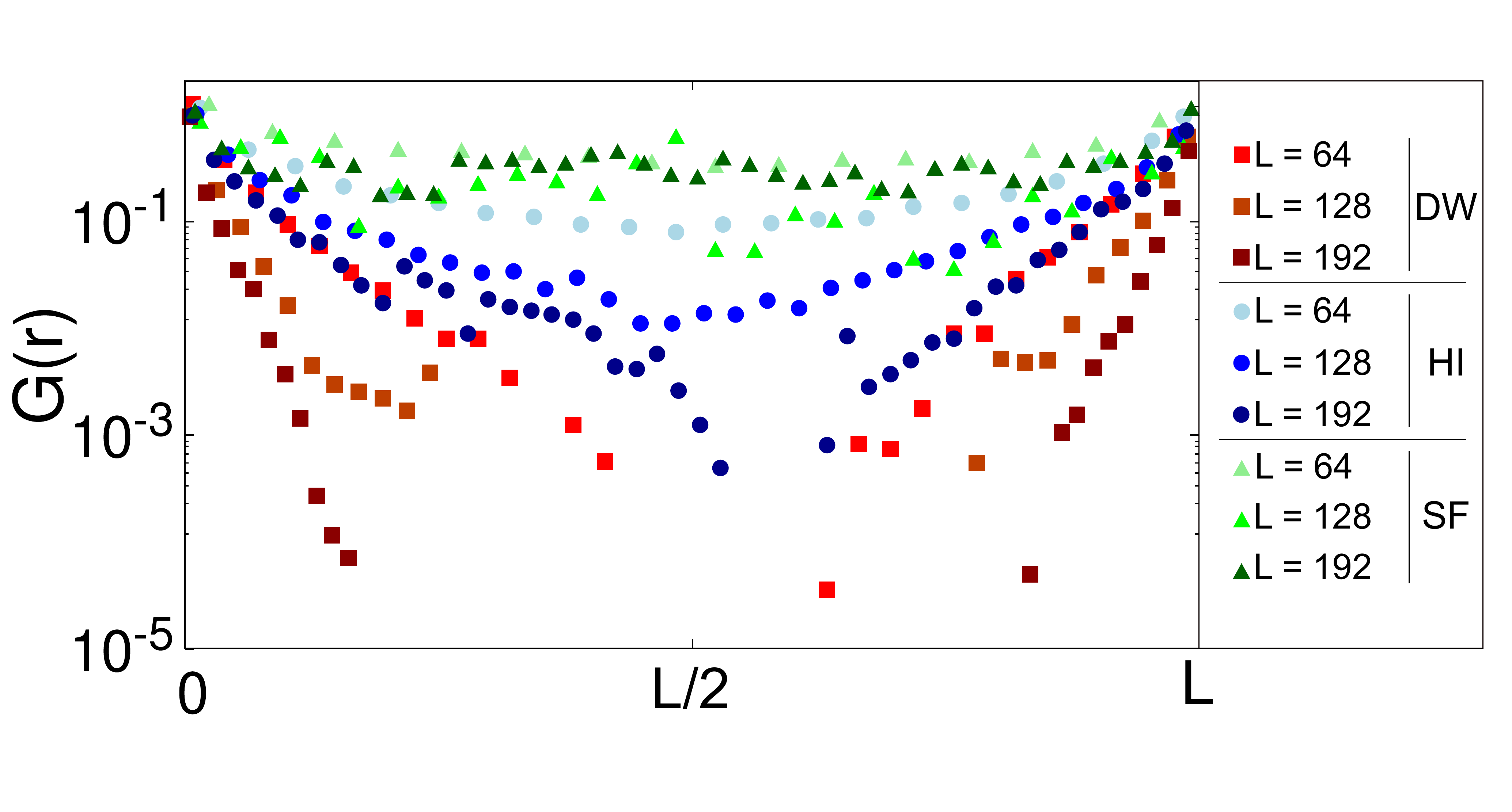}%
\caption{\label{fig:rsGF_DW_HI_SF} Green's function for different phases. Because of periodic boundary conditions, it is $G(r)=G(L-r)$. {\color{red} Red squares}: The DW phase for $t=0.2$. We see an exponential decay (note the logarithmic scale) for all system sizes. $G(r)$ approaches zero for larger distances implying zero winding and absence of superfluid density. {\color{blue}Blue circles}: The HI phase with $t=0.23$. While $G(r)$ for small chain lengths looks similar to $G(r)$ for the SF phase, large sizes display an exponential decay like in the DW phase. {\color{green}Green triangles}: In the SF phase ($t=0.34$) the Green's function is size-independent and $G(r)$ decays algebraically from $r=0$. For larger distances from zero, $G(r)$ flattens due to the periodic boundary conditions.}
\end{figure}

Fig.  \ref{fig:rsGF_DW_HI_SF} depicts the Green’s function $G(r)$ for various chain lengths in the different phases. When the worm moves through the extended configuration space the distance between worm head and tail varies with every Monte Carlo move. In the MI and DW phase, this movement is rather restricted and head and tail stay close to each other, which leads to an exponentially decaying Green's function. In the SF phase bosons become delocalized implying that both worm ends can be arbitrarily far from each other. The Green's function in the SF phase is expected to decay algebraically in one dimension, but in a finite system $G(r)$ has a minimum at $G(L/2)$ due to the periodic boundary conditions, as is visible in Fig.  \ref{fig:rsGF_DW_HI_SF}. In the HI phase $G(r)$ behaves similar to the SF phase for small system sizes, but for larger system sizes the winding of the worm becomes unlikely and - as in the MI and DW phase - the decay of the Green’s function approaches an exponential form.\\
The Fourier transformation of the Green’s function yields the momentum distribution and the $k = 0$ mode gives the condensate fraction, which is experimentally accessible \cite{Koehl2004}. Note that due to the algebraic decay of the Greens function in one dimension the condensate fraction vanishes in the thermodynamic limit, in contrast to the superfluid density.
\subsection{Results for the LR-BHM\label{subsec:ResultsfortheLR-BHM}}
Now set $V=0$ and $U_d=0.6$. For $t=0$ it is straightforward to determine the different phases of eq. (\ref{eq:Hamiltonian}). First, since $2U_d > U$, all phases are DW(X,0) phases with $X$ being any integer number. That means every second site is occupied by $X$ particles, while all other sites are empty. The transition from vacuum to the DW(1,0) phase is at $\mu_{0,1} =-0.3$ and at $\mu_{1,2}=0.1$ the DW(2,0) phase starts. Every DW phase has a width of $\Delta\mu = 0.4$, so the next transitions are at $\mu_{2,3}=0.5$, $\mu_{3,4}=0.9$ and so on.

On the basis of the MF analysis resulting in eq. (\ref{eq:MF_Hamiltonian}) we can obtain a first guess of the approximate shape of the phase diagram of the LR-BHM and introduce a set of rescaled parameters:
$\Utilde = 0.4$, $\mutilde = \mu + 0.3$ and $\Vtilde = 0.3$, thus the ratio $\Vtilde/\Utilde = 0.75$ is identical to the NN case. The rescaled on-site repulsion accounts for the same lobe width $\Delta \mu = \Utilde$ and the rescaled chemical potential is shifted by the equal amount as discussed above for the $t=0$ case. Then the DW(2,0) phase exists in the interval $\mutilde/\Utilde \in [1,2]$, the same range as in the NN-BHM (Fig. \ref{fig:phase_diagram_Ud_0}).
\begin{figure}\centering
\includegraphics[width=\columnwidth]{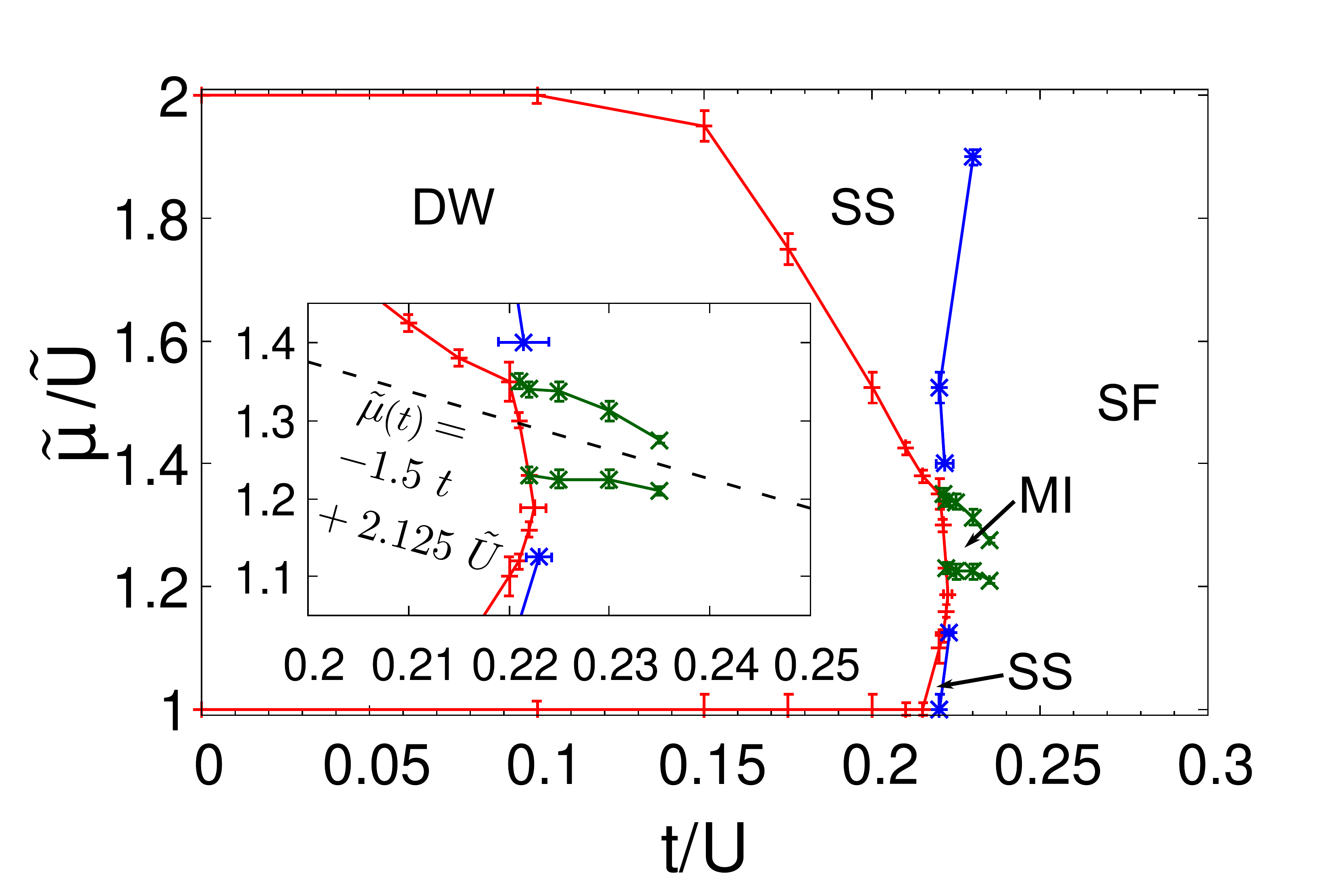}%
\caption{\label{fig:phase_diagram_V_0} Phase diagram of the LR-BHM for $U_d=0.6$. The red line separates the DW(2,0) phase from the SS phase. Through the green line the MI phase transits into the SF phase and the blue line marks the transition from SS to SF phases. The diagram shows the rescaled ratio $\mutilde/\Utilde = (\mu+0.3)/0.4$ for a comparison with the NN-BHM case, as discussed in the mean-field analysis. The inset depicts the tip of the lobe zoomed in and the chemical potential function $\mutilde (t)= -1.5~t+2.125~\Utilde$ used for Fig. \ref{fig:V_0}. }
\end{figure}

Therefore, we present the results for the LR-BHM in the ratio of $\mutilde/\Utilde$ and compare it directly to the NN-BHM case. In Fig. \ref{fig:phase_diagram_V_0} the DW(2,0) lobe is depicted.  In comparison with the phase diagram of the NN-BHM, Fig. \ref{fig:phase_diagram_Ud_0}, we see that no HI phase emerges at the tip of the DW lobe, but a MI phase instead. Otherwise, the DW phase is broader and its tip shifted downwards.

Next, analogous to the NN-BHM case, we keep $\rho=1$ constant by fixing $\mutilde$ to a linear function depended on $t$ and increase $t$ from $0$ to $0.25$.
\begin{figure}\centering
\includegraphics[width=\columnwidth]{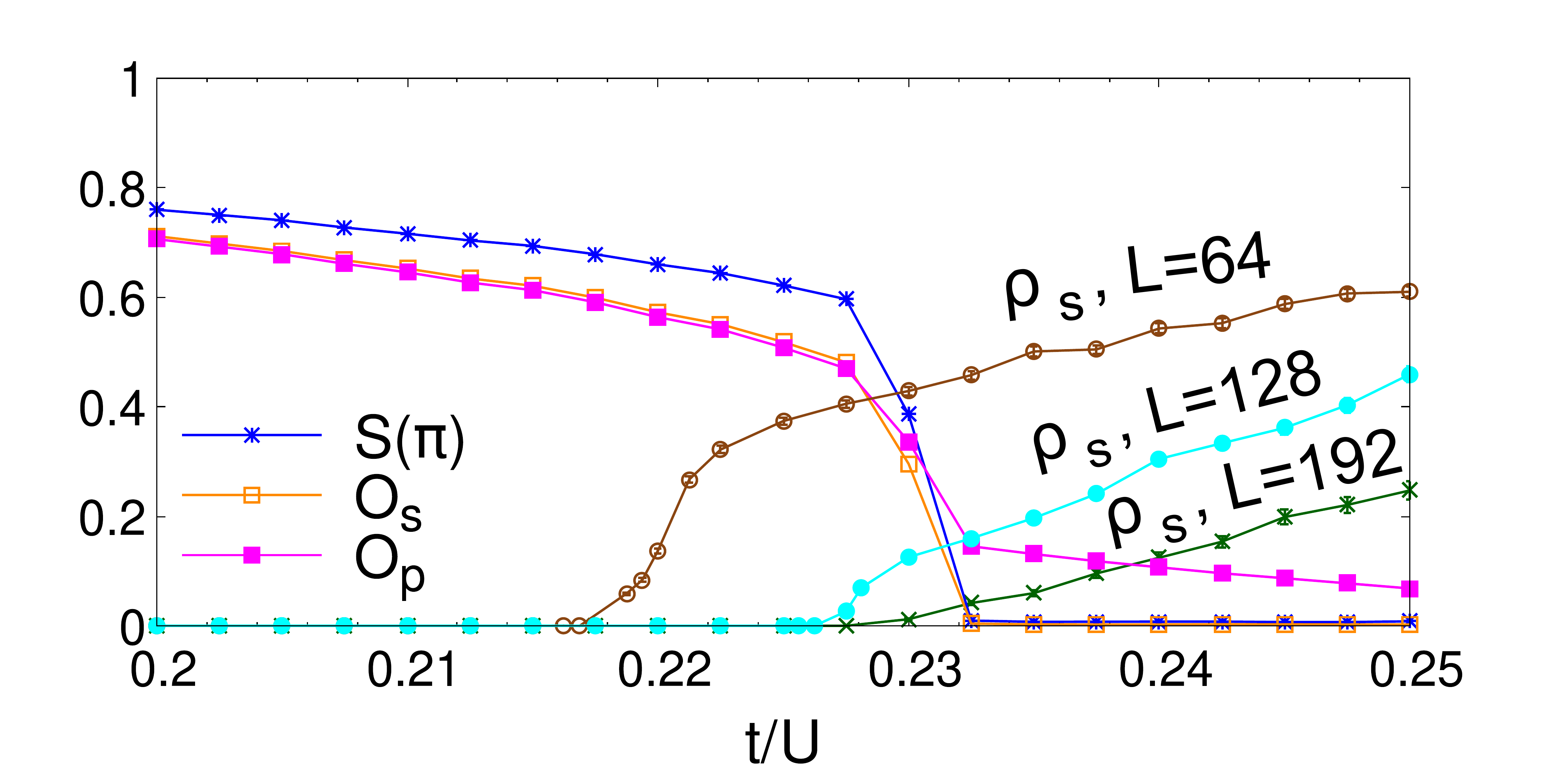}%
\caption{\label{fig:V_0} Order parameters as a function of $t/U$ for $\mutilde (t)= -1.5~t+2.125~\Utilde$ and $U_d=0.6$ in the LR-BHM at $\rho=1$. Depicted are the DW(2,0) phase left to $t=0.23$ and the MI phase right to it. For the DW phase the structure factor, string order parameter and parity order parameter are non-zero, while the superfluid density is zero. In the MI phase the parity order parameter remains greater than zero and the superfluid density increases but approaches zero for larger chain lengths.}
\end{figure}
Our results are shown in Fig. \ref{fig:V_0}. In the DW phase the structure factor, string and parity oder parameters are non-zero, while the superfluid density vanishes. Approaching the transition point at around $t=0.23$ the structure factor and string order parameter drop to zero, while the parity order parameter persists. Also, the superfluid density increases but is strongly dependent on the system size as for larger sizes the superfluid density tends to zero.

The behavior of the $\rho=1$ phase transition in the LR-BHM is similar to the NN-BHM (Fig. \ref{fig:Ud_0}), whereby the MI phase replaces the HI phase. Here, not the string order parameter persists, but the parity.

\begin{figure}\centering
\includegraphics[width=.7\columnwidth]{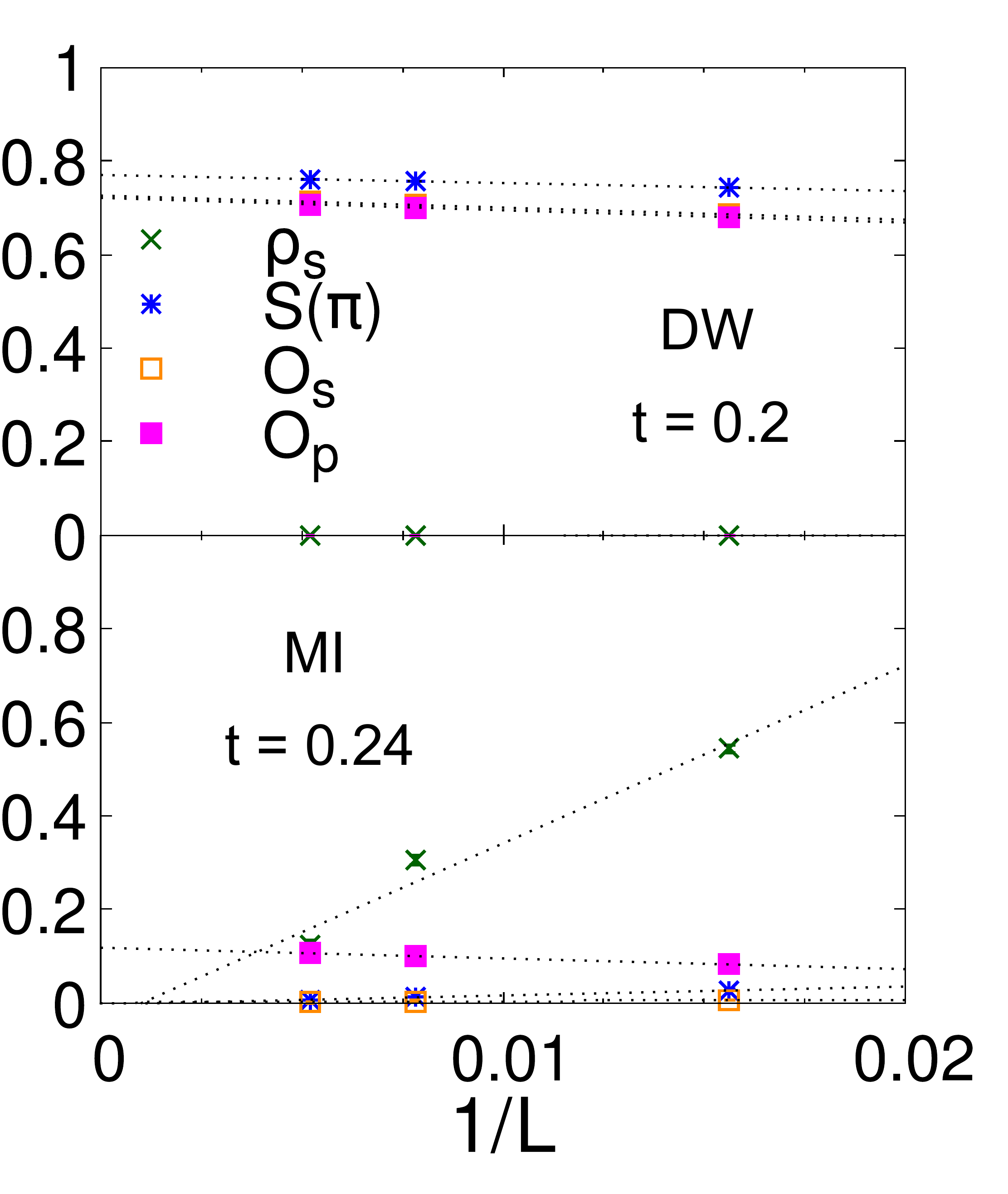}%
\caption{\label{fig:L_scaling_combined_Ud_06} Finite size dependence of different order parameters in the LR-BHM with $\rho=1$. Top: DW phase for $t=0.2$, where structure factor, string order parameter and parity order parameter have finite values, while the superfluid density remains zero. Bottom: MI phase for $t=0.24$, where the structure factor and string order parameter tend to zero, whereas the parity operator has a finite value and the superfluid density vanishes for a large, but finite chain length.}
\end{figure}

We show the finite size scaling of the LR-BHM with commensurate filling in Fig. \ref{fig:L_scaling_combined_Ud_06}. The order parameters in the DW phase behave as in the NN-BHM case. Otherwise, in the MI phase only the parity operator is non-zero, while the superfluid density becomes zero at a large, but finite $L$.

The existence of a MI phase at the tip of the DW phase was not found in two (or more) dimensional systems. The emergence of a MI phase can be understood in the following way:  For the DW(2,0) phase a large long-range interaction exists, preventing site occupation fluctuations from the underlying checkerboard structure. On the other hand in the MI phase for small occupation imbalances, the resulting long-range interaction is proportional to $1/L$, thus negligible for large $L$. This effect is based upon the description of the HI and MI phases as explained above \cite{Berg2008}.

\subsection{Results for the NNLR-BHM\label{subsec:ResultsfortheNNLR-BHM}}
In this section, we analyze nearest-neighbor and long-range interactions simultaneously. As discussed in the subsections before, the HI phase exists in the NN-BHM while it is absent in the LR-BHM. Hence, we start with the NN-BHM and add increasing long-range interactions to see if the HI phase gets destructed.
\begin{figure}\centering
\includegraphics[width=\columnwidth]{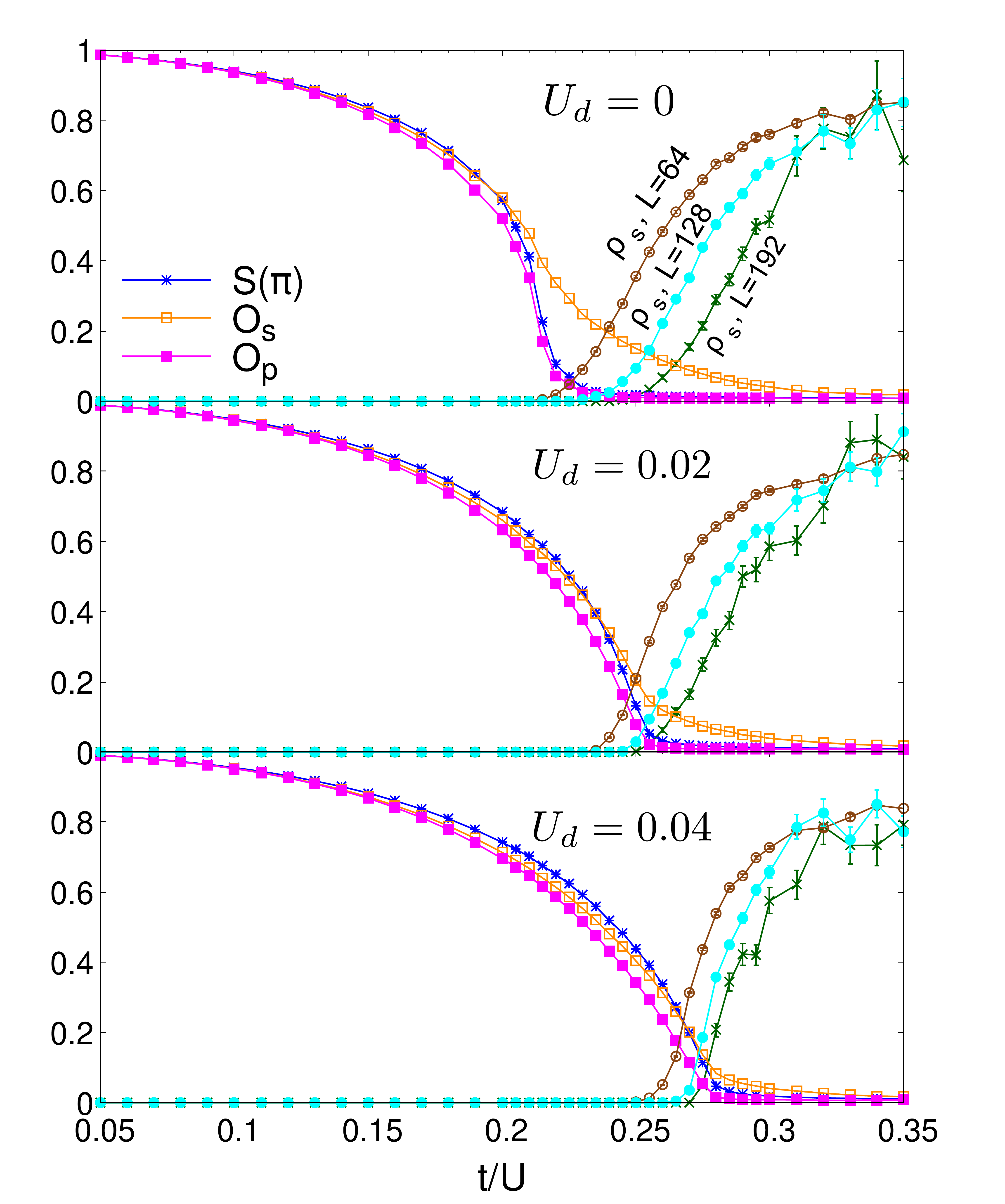}%
\caption{\label{fig:U_d_scaling} Order parameters as a function of $t/U$ for increasing long-range parameters. Top ($U_d = 0$): Same as in Fig. \ref{fig:Ud_0}. Middle ($U_d=0.02$): The DW phase persists up to $t=0.25$ before transiting to the HI phase. Bottom ($U_d=0.04$): Here, the transition takes place at around $t=0.275$. The HI got almost completely occupied by the DW phase.}
\end{figure}

Fig. \ref{fig:U_d_scaling} shows the evolution of the HI phase with the inclusion of LR interaction. Already for LR parameters $U_d$ one order of magnitude smaller than the NN parameter $V$ the HI phase disappears quickly. We see that the phase transition point of $t=0.22$ for the NN-BHM moves to $t=0.25$ for $U_d=0.02$ and to $t=0.275$ for $U_d=0.04$. The phase transition between HI and SF phases is not affected by the increasing LR parameter.

The inclusion of a system wide LR coupling has a strong influence on the DW phase, while it is negligible for the HI and SF phases. Thus, the DW phase expands and supersedes the HI phase. For strong enough LR parameter strength the HI phase disappears completely as expected from the LR-BHM results.

We have not checked the behavior of the MI phase in the LR-BHM when increasing the NN interaction, but we assume this phase to vanish analogously to the HI phase above. Since an additional NN interaction does not increase the energy of the DW(2,0) phase but the energies of the MI and SF phases, we expect the DW(2,0) phase to expand and the MI to shrink for increasing NN interactions.

\section{Conclusions \label{sec:Conclusions}}
In this paper, we investigated the extended 1D Bose-\linebreak Hubbard model with a nearest-neighbor interaction, a \linebreak cavity-mediated long-range interaction and both combined. For the NN-BHM we confirmed earlier results that a Haldane insulator phase exists. For the LR-BHM we found the absence of a HI phase at the tip of the DW lobe and its replacement by a MI phase. The reason is the global long-range interaction preventing site occupation fluctuations in the DW phase while it is possible in the MI phase. For the NNLR-BHM we increased the long-range parameter gradually and 
observed the HI phase to shrink as it is replaced by a growing DW phase. The LR parameter was one order of magnitude smaller than the NN parameter, showing the instability of the HI phase against a global ordering via cavity-mediated interactions.

Furthermore, we conclude that the NN interaction prevents the creation of a MI phase at the tip of the DW lobe due to the commensurate filling of all sites, while the LR interaction suppresses a HI phase at the tip since the specific global order, necessary to form a HI phase, becomes dominated by the long-range ordering effect of the cavity-mediated interactions. Hence, neither HI nor MI phases exist in the NNLR-BHM with strong LR and NN couplings.

\section*{Acknowledgements \label{sec:Acknowledgements}}
We like to thank Benjamin Bogner for his comments on the paper and many beneficial discussions and Chao Zhang for her support with the worm-algorithm and helpful suggestions.
%
%
\section*{Authors contributions}
All the authors were involved in the preparation of the manuscript.
All the authors have read and approved the final manuscript.
%
\bibliographystyle{ieeetr}
\bibliography{bib_epj}

\begin{thebibliography}{10}

\bibitem{Fisher1989}
M.~P.~A. Fisher, P.~B. Weichman, G.~Grinstein, and D.~S. Fisher {\em Phys. Rev.
  B}, vol.~40, pp.~546--570, Jul 1989.

\bibitem{Pollock1987}
E.~L. Pollock and D.~M. Ceperley {\em Phys. Rev. B}, vol.~36, pp.~8343--8352,
  Dec 1987.

\bibitem{Ceperley1995}
D.~M. Ceperley {\em Rev. Mod. Phys.}, vol.~67, pp.~279--355, Apr 1995.

\bibitem{Batrouni1992}
G.~G. Batrouni and R.~T. Scalettar {\em Phys. Rev. B}, vol.~46, pp.~9051--9062,
  Oct 1992.

\bibitem{Batrouni1995}
G.~G. Batrouni, R.~T. Scalettar, G.~T. Zimanyi, and A.~P. Kampf {\em Phys. Rev.
  Lett.}, vol.~74, pp.~2527--2530, Mar 1995.

\bibitem{Prokofev1998}
N.~Prokof'ev, B.~Svistunov, and I.~Tupitsyn {\em Physics Letters A}, vol.~238,
  no.~4, pp.~253 -- 257, 1998.

\bibitem{Prokofev1998_aug}
N.~V. Prokof'ev, B.~V. Svistunov, and I.~S. Tupitsyn {\em Journal of
  Experimental and Theoretical Physics}, vol.~87, pp.~310--321, Aug 1998.

\bibitem{Prokofev2001}
N.~Prokof'ev and B.~Svistunov {\em Phys. Rev. Lett.}, vol.~87, p.~160601, Sep
  2001.

\bibitem{Prokofev2004}
N.~Prokof'ev and B.~Svistunov {\em Phys. Rev. Lett.}, vol.~92, p.~015703, Jan
  2004.

\bibitem{Rousseau2008_may}
V.~G. Rousseau {\em Phys. Rev. E}, vol.~77, p.~056705, May 2008.

\bibitem{Rousseau2008_nov}
V.~G. Rousseau {\em Phys. Rev. E}, vol.~78, p.~056707, Nov 2008.

\bibitem{vanOosten2001}
D.~van Oosten, P.~van~der Straten, and H.~T.~C. Stoof {\em Phys. Rev. A},
  vol.~63, p.~053601, Apr 2001.

\bibitem{Dhar2011}
A.~Dhar, M.~Singh, R.~V. Pai, and B.~P. Das {\em Phys. Rev. A}, vol.~84,
  p.~033631, Sep 2011.

\bibitem{Elstner1999}
N.~Elstner and H.~Monien {\em Phys. Rev. B}, vol.~59, pp.~12184--12187, May
  1999.

\bibitem{Rokhsar1991}
D.~S. Rokhsar and B.~G. Kotliar {\em Phys. Rev. B}, vol.~44, pp.~10328--10332,
  Nov 1991.

\bibitem{Krauth1992}
W.~Krauth, M.~Caffarel, and J.-P. Bouchaud {\em Phys. Rev. B}, vol.~45,
  pp.~3137--3140, Feb 1992.

\bibitem{Kuehner1998}
T.~D. K\"uhner and H.~Monien {\em Phys. Rev. B}, vol.~58, pp.~R14741--R14744,
  Dec 1998.

\bibitem{Jaksch1998}
D.~Jaksch, C.~Bruder, J.~I. Cirac, C.~W. Gardiner, and P.~Zoller {\em Phys.
  Rev. Lett.}, vol.~81, pp.~3108--3111, Oct 1998.

\bibitem{Greiner2002}
M.~Greiner, O.~Mandel, T.~Esslinger, T.~W. H{\"a}nsch, and I.~Bloch {\em
  Nature}, vol.~415, pp.~39--44, Jan 2002.

\bibitem{Kim2004}
E.~Kim and M.~H.~W. Chan {\em Nature}, vol.~427, pp.~225--227, Jan 2004.

\bibitem{Landig2016}
R.~Landig, L.~Hruby, N.~Dogra, M.~Landini, R.~Mottl, T.~Donner, and
  T.~Esslinger {\em Nature}, vol.~532, pp.~476 EP --, Apr 2016.

\bibitem{Hruby2018}
L.~Hruby, N.~Dogra, M.~Landini, T.~Donner, and T.~Esslinger {\em Proceedings of
  the National Academy of Sciences}, vol.~115, no.~13, pp.~3279--3284, 2018.

\bibitem{Batrouni2006}
G.~G. Batrouni, F.~H\'ebert, and R.~T. Scalettar {\em Phys. Rev. Lett.},
  vol.~97, p.~087209, Aug 2006.

\bibitem{Capogrosso-Sansone2007}
B.~Capogrosso-Sansone, N.~V. Prokof'ev, and B.~V. Svistunov {\em Phys. Rev. B},
  vol.~75, p.~134302, Apr 2007.

\bibitem{Capogrosso-Sansone2008}
B.~Capogrosso-Sansone, {\c{S}}.~G. S\"oyler, N.~Prokof'ev, and B.~Svistunov
  {\em Phys. Rev. A}, vol.~77, p.~015602, Jan 2008.

\bibitem{Capogrosso-Sansone2010}
B.~Capogrosso-Sansone, C.~Trefzger, M.~Lewenstein, P.~Zoller, and G.~Pupillo
  {\em Phys. Rev. Lett.}, vol.~104, p.~125301, Mar 2010.

\bibitem{Ohgoe2012_aug}
T.~Ohgoe, T.~Suzuki, and N.~Kawashima {\em Phys. Rev. B}, vol.~86, p.~054520,
  Aug 2012.

\bibitem{Batrouni2013}
G.~G. Batrouni, R.~T. Scalettar, V.~G. Rousseau, and B.~Gr\'emaud {\em Phys.
  Rev. Lett.}, vol.~110, p.~265303, Jun 2013.

\bibitem{Rigol2009}
M.~Rigol, G.~G. Batrouni, V.~G. Rousseau, and R.~T. Scalettar {\em Phys. Rev.
  A}, vol.~79, p.~053605, May 2009.

\bibitem{Kato2009}
Y.~Kato and N.~Kawashima {\em Phys. Rev. E}, vol.~79, p.~021104, Feb 2009.

\bibitem{Greschner2013}
S.~Greschner, L.~Santos, and T.~Vekua {\em Phys. Rev. A}, vol.~87, p.~033609,
  Mar 2013.

\bibitem{Gurarie2009}
V.~Gurarie, L.~Pollet, N.~V. Prokof'ev, B.~V. Svistunov, and M.~Troyer {\em
  Phys. Rev. B}, vol.~80, p.~214519, Dec 2009.

\bibitem{Niederle2016}
A.~E. Niederle, G.~Morigi, and H.~Rieger {\em Phys. Rev. A}, vol.~94,
  p.~033607, Sep 2016.

\bibitem{Zhang2019}
C.~{Zhang} and H.~{Rieger} {\em ArXiv e-prints}, Aug. 2019.

\bibitem{Zhang2015}
C.~Zhang, A.~Safavi-Naini, A.~M. Rey, and B.~Capogrosso-Sansone {\em New
  Journal of Physics}, vol.~17, p.~123014, dec 2015.

\bibitem{Sengupta2005}
P.~Sengupta, L.~P. Pryadko, F.~Alet, M.~Troyer, and G.~Schmid {\em Phys. Rev.
  Lett.}, vol.~94, p.~207202, May 2005.

\bibitem{Iskin2011}
M.~Iskin {\em Phys. Rev. A}, vol.~83, p.~051606, May 2011.

\bibitem{Kimura2011}
T.~Kimura {\em Phys. Rev. A}, vol.~84, p.~063630, Dec 2011.

\bibitem{Rossini2012}
D.~Rossini and R.~Fazio {\em New Journal of Physics}, vol.~14, p.~065012, jun
  2012.

\bibitem{Kawaki2017}
K.~Kawaki, Y.~Kuno, and I.~Ichinose {\em Phys. Rev. B}, vol.~95, p.~195101, May
  2017.

\bibitem{Herbert2001}
F.~H\'ebert, G.~G. Batrouni, R.~T. Scalettar, G.~Schmid, M.~Troyer, and
  A.~Dorneich {\em Phys. Rev. B}, vol.~65, p.~014513, Dec 2001.

\bibitem{Schmid2004}
G.~Schmid and M.~Troyer {\em Phys. Rev. Lett.}, vol.~93, p.~067003, Aug 2004.

\bibitem{DallaTorre2006}
E.~G. Dalla~Torre, E.~Berg, and E.~Altman {\em Phys. Rev. Lett.}, vol.~97,
  p.~260401, Dec 2006.

\bibitem{Chen2008}
Y.-C. Chen, R.~G. Melko, S.~Wessel, and Y.-J. Kao {\em Phys. Rev. B}, vol.~77,
  p.~014524, Jan 2008.

\bibitem{Dogra2016}
N.~Dogra, F.~Brennecke, S.~D. Huber, and T.~Donner {\em Phys. Rev. A}, vol.~94,
  p.~023632, Aug 2016.

\bibitem{Flottat2017}
T.~Flottat, L.~d.~F. de~Parny, F.~H\'ebert, V.~G. Rousseau, and G.~G. Batrouni
  {\em Phys. Rev. B}, vol.~95, p.~144501, Apr 2017.

\bibitem{Bogner2019}
{Bogner, Benjamin}, {De Daniloff, Cl\'ement}, and {Rieger, Heiko} {\em Eur.
  Phys. J. B}, vol.~92, no.~5, p.~111, 2019.

\bibitem{Berg2008}
E.~Berg, E.~G. Dalla~Torre, T.~Giamarchi, and E.~Altman {\em Phys. Rev. B},
  vol.~77, p.~245119, Jun 2008.

\bibitem{Haldane1983}
F.~Haldane {\em Physics Letters A}, vol.~93, no.~9, pp.~464 -- 468, 1983.

\bibitem{Haldane1983_2}
F.~D.~M. Haldane {\em Phys. Rev. Lett.}, vol.~50, pp.~1153--1156, Apr 1983.

\bibitem{Pollet2007}
L.~Pollet, K.~V. Houcke, and S.~M. Rombouts {\em Journal of Computational
  Physics}, vol.~225, no.~2, pp.~2249 -- 2266, 2007.

\bibitem{Kennedy1992}
T.~Kennedy and H.~Tasaki {\em Phys. Rev. B}, vol.~45, pp.~304--307, Jan 1992.

\bibitem{Koehl2004}
M.~Köhl, T.~Stöferle, H.~Moritz, C.~Schori, and T.~Esslinger {\em Applied
  Physics B}, vol.~79, Dez 2004.

\end{thebibliography}
%

\end{document}